\documentclass[aps,prb,preprintnumbers,footinbib,floatfix,superscriptaddress]{revtex4}
\usepackage{graphicx}
\usepackage{bm}
\usepackage{amsfonts}
\usepackage{amsmath}
\usepackage[percent]{overpic}
\usepackage{xcolor}
\usepackage[colorlinks=true,citecolor=blue]{hyperref}

\hypersetup{colorlinks=true,citecolor=blue,linkcolor=red,urlcolor=blue}

\begin{document}

\title{Zero point motion and direct/indirect bandgap crossover in layered
  transition-metal dichalcogenides}
\date{\today}

\author{Luciano Ortenzi}
\affiliation{Istituto dei Sistemi Complessi, CNR, 00185 Roma, Italy}
\affiliation{Dipartimento di Fisica, Universit\`a La Sapienza, P.le A. Moro 2,
00185 Roma, Italy }

\author{Luciano Pietronero}
\affiliation{Dipartimento di Fisica, Universit\`a La Sapienza, P.le A. Moro 2,
00185 Roma, Italy }
\affiliation{Istituto dei Sistemi Complessi, CNR, 00185 Roma, Italy}

\author{Emmanuele Cappelluti}
\affiliation{Istituto di Struttura della Materia, CNR,
Division of Ultrafast Processes in Materials (FLASHit),
34149 Trieste, Italy}

\begin{abstract}
Two-dimensional transition-metal dichalcogendes $MX_2$
(es. MoS$_2$, WS$_2$, MoSe$_2$, \ldots) are among the most
promising materials for bandgap engineering.
Widely studied in these compounds,
by means of ab-initio techniques,
is the possibility of tuning
the direct-indirect gap character by means of in-plane strain.
In such kind of calculations however the lattice degrees of freedom
are assumed to be classical and frozen.
In this paper we investigate in details the dependence of the bandgap
character (direct vs. indirect) on the out-of-plane 
distance $h$ between the two chalcogen planes in each $MX_2$ unit.
Using DFT calculations, we show that the bandgap character is indeed highly sensitive
on the parameter $h$, in monolayer as well as in bilayer and bulk compounds,
permitting for instance the switching
from indirect to direct gap and from indirect to direct gap in monolayer systems.
This scenario is furthermore analyzed in the presence of quantum lattice
fluctuation induced by the zero-point motion.
On the basis of a quantum analysis, we argue that the direct-indirect
bandgap transitions induced by the out-of-plane strain as well by the
in-plane strain can be regarded more as continuous crossovers rather
than as real sharp transitions. The consequences
on the physical observables are discussed.
\end{abstract}
\maketitle

\section{Introduction}

The isolation of graphene, in 2004,\cite{novoselov1,novoselov2}
has opened the doors
for intensive research on two-dimensional materials, i.e. layered materials that
can be grown/exfoliated to atomical thickness.
Within this context,
layered transition-metal dichalcogenides (TMDs) $MX_2$
($M=$Mo, W; $X=$S, Se) appear as the most promising
compounds for future technological applications.
The presence of many degrees of freedom (charge, spin, valley, layer,
lattice, ldots), strongly entangled each other,
provides a fruitful playground for designing devices where
electronic/optical/magnetic/transport
properties can be tuned in a controlled and reversible way
by external conditions, e.g. magnetic/electric fields,
pressure, temperature, strain. 
\cite{radi,zeng2,mak2,wang2,cao,zeng3,wu2,wang3,terrones,xu,ganatra,andres,review-2Dmat,review-theory,seo-kim}
A striking difference of these materials with respect to graphene is that
semiconducting TMDs present an intrinsic bandgap,
\cite{bromley,mattheis,lebegue,mak2010}
more suitable thus for technological applications\cite {reviews-tech}
such as flexible electronics,\cite{kis,fiori,desai}
nanophotonics,\cite{tsai, kis2, koppens,nanoP}
(photo)-catalysis,\cite{benck,baek,zeng,pumera,catalysis}
optoelectronics,\cite{eda}
etc\ldots
 
Interestingly, not only the size of the bandgap, but also the
character (direct/indirect) can be relatively easily controlled.
In MoS$_2$, for instance, the bandgap energy $E_{\rm g}$
changes from 1.29 eV
in the bulk compounds to 1.90 eV in the monolayer.
At the same time, the bandgap evolves from
an indirect one for $N \ge 2$ ($N$ being the number of layers)
to a direct one for the monolayer compounds.\cite{mak2010,splendiani2010}

\begin{figure}[t]
\includegraphics[width=12cm]{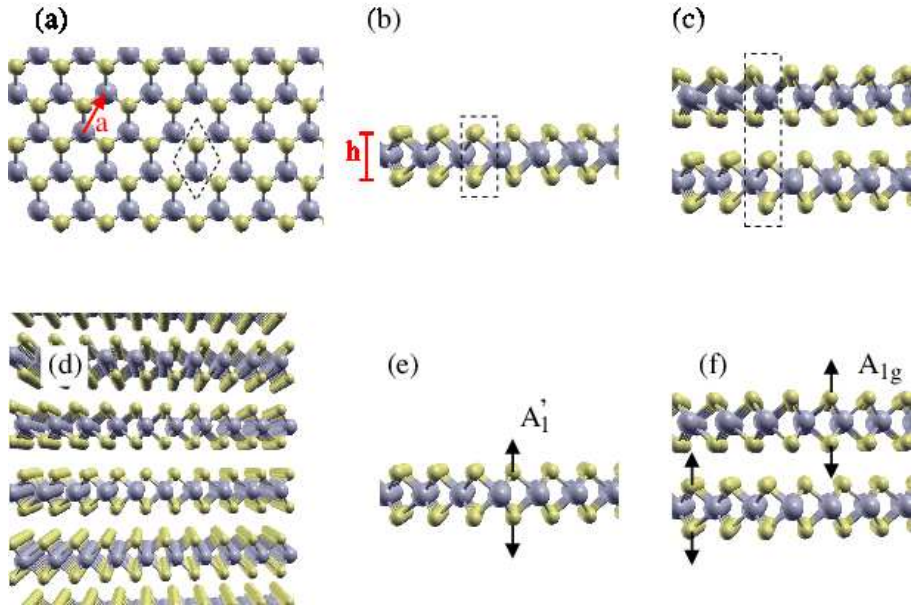}
\caption{(a) Top view and (b) side view of monolayer transition-metal
dichalcogenides $MX_2$. The unit cell (marked by dashed lines)
contain two $X$ atoms and one
$M$ atom. In the top view, the bottom $X$ atom is hidden by the top
one.
The structural parameters $a$ and $h$ are here also defined.
(c) Side view of bilayer $MX_2$ in the 2H-$MX_2$ structure.
The unit cell
is here specified.
(d) Side view of bulk $MX_2$ in the 2H-$MX_2$ structure. (e)-(f)
Lattice displacements of the $A_1^\prime$ and $A_{1g}$ modes
in the single layer and bilayer/bulk structure, respectively.
}
\label{f-structure}
\end{figure}

Thanks to the strong entanglement between electronic and lattice degrees
of freedom, mechanical deformations (strain, pressure, \ldots) of TMDs are among the best candidates for controlling and tuning
the electronic properties.
For instance, within this context,
in monolayer MoS$_2$ density-functional-theory (DFT) calculations have predicted
a change from direct to indirect for strain $\gtrsim 1-4 \%$,
both for compressive and tensile in-plane strain.\cite{scalise1,johari,lu,yun,peelaers,shi,horzum,ghorbani,zhang,scalise2,guzman,wang,roldan,Ye,Gong}
Most of the ab-initio works in this field assume classical
(and static)
lattice coordinates, where the electronic bands
are evaluated in the perfect crystal structure (see
Fig. \ref{f-structure} for a sketch of the crystal structure).
Most of the papers, also, focus their interest on the role
of the in-plane strain, whereas the internal coordinates within a
single sandwich $MX_2$,
i.e. the ``thickness'' of the $MX_2$,
characterized by the vertical distance $h$ between the chalcogen planes,
is assumed to be unchanged,\cite{johari,lu,ghorbani,scalise2,guzman} or
it is optimized upon lattice relaxation.\cite{scalise1,shi,horzum,zhang,scalise2,wang}
A couple of recent papers have on the other hand
focused on the role of the vertical $X$-$X$ distance.\cite{alvarez,angel-morales}
Using first-principle calculations, they
fix the in-plane lattice parameter and they investigate
the change in the electronic structure under uniaxial vertical
compression, which in the monolayer case it is directly reflected
in the chalcogen distance $h$.
In this way they pointed out how out-of-plane compression of monolayer
sample can be an efficient way for lifting the valence band-edge at
$\Gamma$, and driving thus the system towards a direct/indirect gap
transition with $\Gamma$K indirect-gap character.
This is due to the high sensitivity of the band structure of TMDs on
the vertical $X$-$X$ distance $h$ (Fig. \ref{f-structure}).
Such direct/indirect gap transition driven by out-of-plane compression
is of course less efficient in bilayer-multilayer compounds
where the most of the out-of-plane compression is employed
in reducing the interlayer distance between $MX_2$ sheets.
Note that the vertical $X$-$X$ distance $h$ was there considered
as a classic (frozen) parameter, and in this perspective
only reductions of $h$ were considered, as a result of pressure.

In the present paper 
we show how the pointed out
dependence of the electronic structure on the $X$-$X$ distance $h$ is
much more relevant then in
Refs. \onlinecite{alvarez,angel-morales}.
In particular we show how
the intrinsic quantum lattice fluctuations associated
with the parameter $h$ can probe at a dynamical level
a much larger range of
both direct/indirect bandgap configurations than
pure ``classical'' compression,
so that
the direct/indirect bandgap character
cannot be captured by the analysis of the band-structure of the
perfect crystal (or of any ``effective'' band structure).
In this situation
the basic assumptions underlying the adiabatic
Born-Oppenheimer principle are no more fulfilled
and the very idea of a well-defined direct/indirect bandgap character
is questionable.
We show how this challenging scenario,
which is more striking for monolayer compunds
where a dynamical direct-gap to indirect gap
can be driven, is also relevant in
bilayer/multilayer compounds where a dynamical transition
between two different kinds of indirect bandgap configurations,
with different physical properties, can be induced.

The paper is organized as follows:
in Sec. \ref{mos2.1L} we consider monolayer MoS$_2$
as textbook example to show the relevance of
the zero point motion quantum lattice fluctuations
and of the corresponding dynamically-induced direct/indirect-gap
crossover;
in Sec. \ref{III} the analysis is generalized in the wider context
of monolayer, bilayer and bulk  MoS$_2$, MoSe$_2$, WS$_2$ and WSe$_2$
investigating different kinds of indirect ($\Gamma$K$\leftrightarrow$KQ)
bandgap transitions;
in Sec. \ref{IV} we revise the current understanding
of possible direct/indirect bandgap transitions
in monolayer MoS$_2$ driven by
 in-plane strain in the light of the scenario
prompted by the analysis of the lattice
quantum fluctuations on strained  MoS$_2$ monolayer;
finally, in Sec.\ref{s:bo}
the consequences of the above scenario are discussed.

\section{Direct to indirect bandgap crossover
induced by quantum
  fluctuations in single-layer MoS$_2$}
\label{mos2.1L}

In this Section we investigate the role of the interplane distance $h$
in determining the direct/indirect character of the band-structure
in single-layer MoS$_2$, with a particular regard about the effects of
quantum lattice fluctuations.
This analysis in single-layer MoS$_2$, besides to be highly interesting
by itself, will be used as a template to introduce the relevant concepts
for the further investigation of generic transition-metal
dichalcogenides $MX_2$ ($M$=Mo, W; $X$=S, Se).

\subsection{Frozen phonon calculations}

\label{s-frozen}

In the following, unless specified,
we employ DFT calculations
using the generalized gradient approximation (GGA) with the
linear augmented plane wave (LAPW) method as implemented
in the WIEN2 code.\cite{wien2k,DFT:PBE} Up to 1300
{\bf k} points were used in the self-consistent calculations
with an LAPW basis defined by  the  cutoff $R_SK_{max} =9$. 
The lattice parameters, for bulk MoS$_2$ were taken from
Ref. \onlinecite{brumme}
(i.e. the in-plane lattice constant $a=3.197$ \AA),
and single layer MoS$_2$ was simulated by increasing
the spacing $c'$ between layers until effective decoupling is achieved.
The value of the interplane S-S distance $h$
was taken initially to be the experimental one, $h_{\rm exp}=3.172$ \AA.

\begin{figure}[t]
\includegraphics[height=6.3cm,clip=]{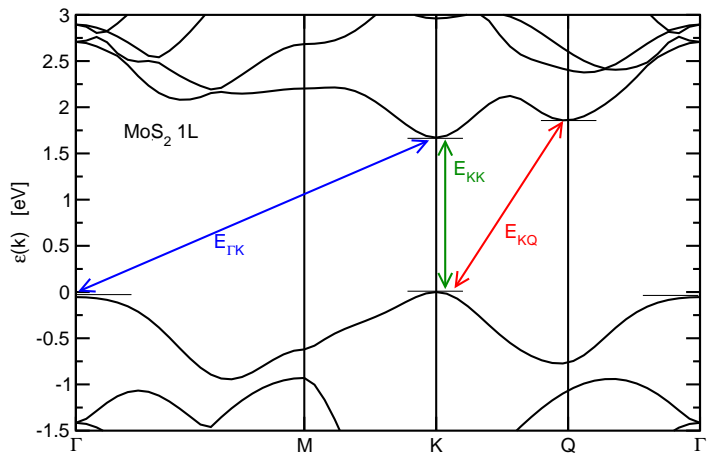}
\caption{
Characteristic band structure of monolayer MoS$_2$, here evaluated
within GGA calculations, with
$a=3.197$ \AA\, and $h_{\rm exp}=3.172$ \AA.
The electronic
dispersion is characterized by three gaps: the direct one $E_{\rm KK}$
at the K point; the indirect one $E_{\Gamma \rm K}$ between the
band edge of the valence band at the $\Gamma$ point and the band edge
of the conduction band at the K point; 
the indirect one $E_{\rm KQ}$ between the
band edge of the valence band at the K point and the band edge
of the conduction band at the Q point.
}
\label{f-bandsMoS2}
\end{figure}
The band structure of single-layer MoS$_2$ crystal,
using these lattice parameters as representative,
has been reported in uncountable papers,
and it is established to have a direct gap $E_{\rm KK}$ at the K point of the
Brillouin zone, as shown in Fig. \ref{f-bandsMoS2}.
A secondary minimum in the conduction band is also present
close the Q point (halfway between K and $\Gamma$) with
a energy difference $\Delta E_c(\mbox{Q})$ of the order of few tenths of meV.
The valence band is also characterized by a secondary maximum at the
$\Gamma$ point, with a slightly lower energy $\Delta E_v(\Gamma)$
than the top-band at the K point.
The precise evaluation of $\Delta E_c(\mbox{Q})$, $\Delta E_v(\Gamma)$
depends on computational details, as the use of the experimental or relaxed lattice
coordinates, the DFT functional used (GGA vs LDA),
the inclusion of many-body effects in a $GW$ scheme, etc.

The high sensitivity of the relative energy differences
$\Delta E_c(\mbox{Q})$, $\Delta E_v(\Gamma)$
on the lattice coordinates provides a powerful tool to
search for a lattice-driven direct/indirect gap transition.
Along this line, the actual possibility of a direct/indirect gap
transition in the band structure, induced by in-plane (uniaxial or biaxial) strain,
has been also extensively investigated by means of ab-initio techniques.
It is widely accepted that a direct/indirect bandgap transition can
occur for biaxial strains of about 2 \%, driving the compounds, for
tensile strain, towards
an indirect gap $E_{\Gamma{\rm K}}$ between the $\Gamma$ point of the
valence band and the K point of the conduction band, 
and towards an indirect gap $E_{\rm KQ}$ between the K point of the
valence band and the Q point of the conduction band, for  compressive
strain.
In most of the papers, the lattice coordinates have been considered
as classical (frozen) variables, in the absence as well as in the
presence of strain.
In many cases, the $h$ coordinate
has been relaxed, so that an in-plane strain induces a change of the
static $h$,
according with the relative Poisson's ratio.

An important step further in this analysis is the disentangling the role
of the {\em in-plane} strain (namely, stretching/compression of the
lattice parameter $a$) from the role of the {\em out-of-plane} strain
(namely, stretching/compression of the structural parameter $h$).\cite{alvarez,angel-morales}
\begin{figure}[t]
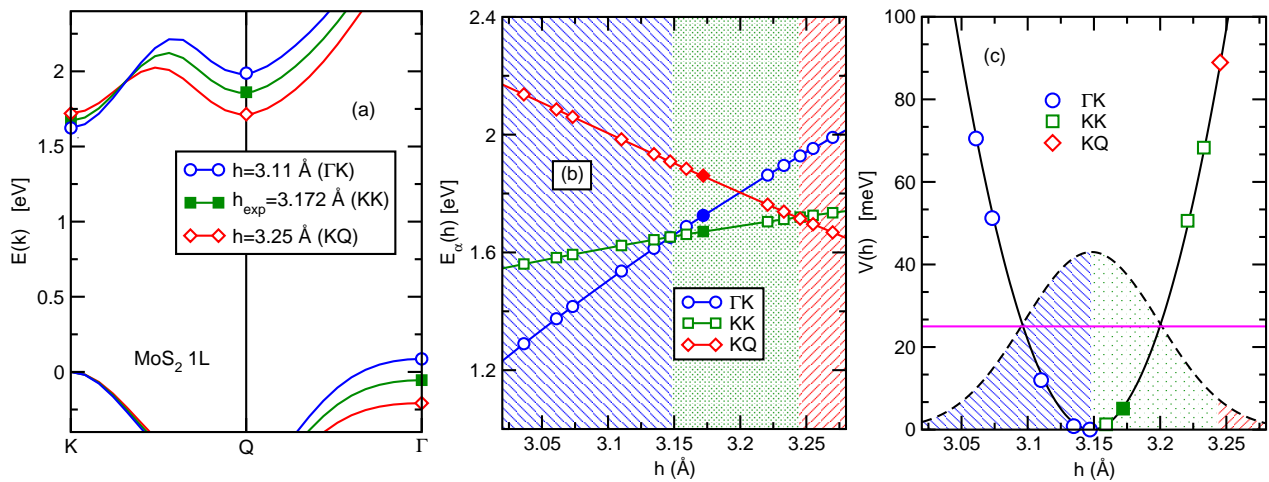

\includegraphics[height=6.3cm,clip=]{f-bands_MoS2_1L_vs_h_2.eps}
\includegraphics[height=6.3cm,clip=]{f-gap_MoS2_1L_vs_h_2.eps}
\includegraphics[height=6.3cm,clip=]{f-frozen_MoS2_1L_vs_h.eps}
\caption{Direct/indirect bandgap crossover for monolayer MoS$_2$
as a function of the
interplane $S$-$S$ structural parameter $h$:
(a) evolution of the band structure for three representative values of $h$.
(b) evolution of the bandgaps $E_\alpha(h)$ as function of $h$.
An indirect gap $E_{\Gamma\rm K}$ is predicted for $h \le 3.148$ \AA;
a direct gap $E_{\rm KK}$ is predicted for $3.148$ \AA $\le h \le 3.244$ \AA;
and an indirect gap $E_{\rm KQ}$ is predicted for $h \ge 3.244$ \AA.
(c) Frozen phonon energy $V(h)$ as a function of $h$ (open symbols).
The shape and color of the symbols  show the character of the gap as
evaluated by panel b. The horizontal line denotes the zero point
energy $\hbar \omega_{A_1^\prime}$, whose intercept  with $V(h)$ gives
the zero point motion lattice displacement. The dashed gaussian-like line is the
probability distribution $P(h)$ defined in Eq. (\ref{Ph}), weighting
the available phase space for each direct/indirect gap case.
}
\label{f-bandsMoS2_h}
\end{figure}
In Fig. \ref{f-bandsMoS2_h}a we show the band structure of monolayer
MoS$_2$ along the cuts K-Q-$\Gamma$ for few representative values
of the inter-plane parameter $h$ 
at fixed value of the in-plane lattice constant $a=3.197$ \AA.
Unlike Refs. \onlinecite{alvarez,angel-morales}, herein we consider both tensile and compressive strain.
We can see that relatively small changes in  $h$ can have drastic
effects
on the band structure and on the direct/indirect character of the
bandgap.
The experimental value $h=3.172$ \AA\, has been marked with filled
squares in Fig. \ref{f-bandsMoS2_h}a
and it shows the typical direct gap at the K point widely reported
in literature.
With respect to this case, we find that
further reduction of $h$, for vertical compression, would lead however
to a indirect gap $E_{\Gamma{\rm K}}$, whereas further increase of $h$, for
vertical tensile elongation, would result in a indirect gap 
$E_{\rm KQ}$.
The plot of the different gaps $E_\alpha$ ($\alpha=\Gamma$K, KK, KQ) 
as functions of $h$ is shown in Fig. \ref{f-bandsMoS2_h}b,
pointing out, in this regime, a linear behavior of the gaps $E_\alpha$
with $h$.
This allows us to estimate a transition between indirect ($\Gamma$K)
to direct (KK) gap at $h_{c1}=3.148$ \AA,
and a transition between direct (KK) to indirect (KQ) gap at
$h_{c2}=3.244$ \AA.
A change of about 3 \% of the interplane distance $h$ appears thus
sufficient to change radically the character of the bandgap from
indirect ($\Gamma$K) to direct (KK) and to indirect (KQ).

The inclusion of the spin-orbit coupling would change
quantitatively but not qualitatively this scenario.
Spin-orbit is indeed known to
induce a band spin-splitting which is largest at the K point
of the valence band ($~ \pm 75$ meV in MoS$_2$)
and at the Q point of the conduction band,
while is relatively smaller
for all the other band edges.\cite{noi2}
Inclusion of the spin-orbit coupling would thus mainly lower
the KK and KQ bandgaps in Fig. \ref{f-bandsMoS2_h}b
of about $\sim 100$ meV, slightly
changing the crossing points $h_{c1}$ and $h_{c2}$.
The change in $h_{c1}$, $h_{c2}$
is expected to be more sizable for WS$_2$ where
the spin-orbit coupling is larger.
As we are going to see in the next Section, however,
the precise determination of $h_{c1}$ and $h_{c2}$
is of secondary relevance since quantum lattice fluctuations
can intrinsically span a large space of $h$.
In addition, the effects of the spin-orbit coupling
are expected to be even less relevant in multilayer ($N \ge 2$)
systems, where the band splitting induced by the spin-orbit coupling
is largely overcome by the concomitant splitting due to the
interlayer coupling.

\subsection{Zero point motion quantum lattice fluctuations}

The high sensitivity of the band structure on the interplane distance
$h$ calls for a more precise assessment of the physical relevant range
of $h$ for real materials.
In order to address this point, we show in Fig. \ref{f-bandsMoS2_h}c
the total energy $V(h)$ of monolayer MoS$_2$ as a function
of the parameter $h$.
The different colored symbols mark the direct/indirect character of
the bandgap, whose band structure is shown in the panel (b).
The minimum of $V(h)$ corresponds to the 
optimized (relaxed) interplane S-S distance, which we found
$h_{\rm cl}=3.148$ \AA, at the very verge of the transition between
a direct and indirect bandgap character.
Such value $h_{\rm cl}=3.148$ \AA\, represents thus the
vertical interplane S-S distance at the classical level,
i.e. in the frozen perfect crystal structure, corresponding
to a gap at a classical level $E_{\rm cl}=1.653$ eV with direct KK character.

The relevance of indirect gap states even in single layer MoS$_2$
appears however striking when
intrinsic quantum lattice fluctuations are taken into account.

To address this issue in more details, we notice that the interplane S-S
distance $h$ is directly associated
with the phonon $A_1^\prime$ whose lattice displacements are shown in
Fig. \ref{f-structure}e.\cite{zhang.phonon,phonon.exp}
The total energy $V(h)$ reported in Fig. \ref{f-bandsMoS2_h}c
can be viewed thus precisely as the frozen phonon
$A_1^\prime$, representing the energy profile of the system as a function
of the static phonon coordinate $h$.
Quantum lattice fluctuations can be formally evaluated by
fitting $V(h)$ and
solving the associated quantum Schroedinger's equation for the lattice
displacement.\cite{bcp,cp}
This task can be further simplified by noticing that
the frozen phonon energy potential obeys
to a perfect parabolic profile $V(h)=a_2(h-h_{\rm cl})^2$,
with $a_2= 4.68$ eV/\AA$^2$, pointing
out to the absence of anharmonic effects.
The zero point motion mean square value of the quantum lattice
fluctuations,
$h_{\rm ZPM}$, can be analytically obtained by the condition
$V(h_{\rm cl}\pm h_{\rm ZPM})=N_{\rm S}\hbar \omega_{A_1^\prime}/4$,
as graphically shown in Fig. \ref{f-bandsMoS2_h}c.
Here $N_{\rm S}=2$ is the number of sulphur atoms for unit cell,
and $\hbar \omega_{A_1^\prime}$ the energy of the $A_1^\prime$ phonon.
From this simple analysis we get $h_{\rm ZPM}=0.051$ \AA.
Such estimate is corroborated by a direct 
solution of the Schroedinger's equation,
that permits us to
evaluate the ground-state wave-function $\Psi_{\rm G}(h)$
of the $A_1^\prime$ phonon,
and hence the probability distribution function $P(h)=|\Psi_{\rm G}(h)|^2$.
Given the harmonic character,
$P(h)$ can be written as:
\begin{eqnarray}
P(h)
&=&
\frac{1}{\sqrt{2\pi h_{\rm ZPM}^2}}
\exp\left[
-\frac{(h-h_{\rm cl})^2}{h_{\rm ZPM}^2}
\right].
\label{Ph}
\end{eqnarray}
The function $P(h)$ represents the (ground-state) probability for the compound
to have S-S inter-plane distance $h$.
The plot of $P(h)$ is also shown in Fig. \ref{f-bandsMoS2_h}c,
showing that the direct bandgap character
found by static ab-initio calculations can be
poorly representative of the rich phase space of this compound,
where zero point motion
quantum lattice fluctuations are expected to span dynamically
regions with a direct gap as well as regions with a indirect gap,
both with $\Gamma$-K and K-Q character.

To gain a deeper insight about this issue
we can thus define,
within a semi-classical framework,
a probability distribution for the gap function,
$P_E(E)=P[E(h)]$, where there is a one-to-one correspondence between the variable $h$
and the bandgap value $E_\alpha$ with character $\alpha$.
In the same spirit, we can thus introduce two compact parameters, namely:
\begin{eqnarray}
W_\alpha
&=&
\int_{E(h) \in E_\alpha} dE P_E(E)
,
\label{weights}
\end{eqnarray}
and
\begin{eqnarray}
E_\alpha
&=&
\int_{E(h) \in E_\alpha} dE E P_E(E)
.
\label{averages}
\end{eqnarray}

The parameter $W_\alpha$ represents
the (ground-state) probabilities
to have direct or indirect gap with character $\alpha=\Gamma$K, KK, KQ,
whereas $E_\alpha$ represents  {\em average} bandgap $\bar{E}_\alpha$
for each direct/indirect gap.

For monolayer MoS$_2$ we find $W_{\Gamma{\rm K}}=0.50$,
$\bar{E}_{\Gamma{\rm K}}=1.502$ eV,
$W_{\rm KK}=0.47$, $\bar{E}_{\rm KK}=1.671$ eV, 
and $W_{\rm KQ}=0.03$, $\bar{E}_{\rm KQ}=1.681$ eV.
This shows that, although classical bandstructure calculations with
frozen lattice dynamics would predict a large direct bandgap with
$E_{\rm KK}\approx 1.653-1.670$ eV (depending on the optimized or
experimental value of $h$), 
quantum lattice fluctuations would suggest rather a dominance of an
indirect $\Gamma$K gap with $\bar{E}_{\Gamma{\rm K}}\approx 1.50$ eV,
besides an additional small component with indirect KQ gap with 
$\bar{E}_{\rm KQ}=1.68$ eV.

This situation is radically unconventional compared with the physics
of standard semiconductors, where the quantum lattice fluctuations
do not affect at such extent the value of the bandgap, nor are able to switch
the direct/indirect character of the bandgap.
On the contrary,
at a qualitative level,
it appears in these compounds that electronic configurations
with indirect and direct gap can be dynamically probed
as a result of quantum lattice fluctuations.
In this scenario it appears thus impossible to disentangle
the quantum dynamics of the lattice degrees of freedom
from the electronic excitations, leading to
the breakdown of the Born-Oppenheimer 
adiabatic assumption, which is on the base of the most of the concepts of
solid state physics.
A rigorous theory taking into account nonadiabatic processes induced
by the breakdown of the Born-Oppenheimer principle
is a formidable task that is at the moment lacking in literature.
In Sec. \ref{s:bo} we discuss in more details a possible approach
to tackle this scenario, compared with the relevant studies in literature,
and possible physical consequences of this scenario.

\section{Quantum lattice fluctuations and bandgaps in layered
transition-metal dichalcogenides}
\label{III}
In the previous Section, we have considered monolayer MoS$_2$ as
representative case of layered transition-metal semiconductor
dichalcogenides, with particular focus on the relevance of the interplane S-S distance $h$ in regards to the bandgap character.
This analysis has permitted us to point out a possible crossover,
driven by the quantum lattice fluctuations,
between direct and indirect gap. To this aim we have introduced useful quantities
as the weights $W_\alpha$ and the average bandgaps $\bar{E}_\alpha$
that describe in a synthetic way the relevance
of the different electronic band-structures as probed by the lattice
quantum fluctuations.

The relevance of these effects is now investigated in a systematic way
in the broad family of multilayer TMDs $MX_2$,
were $M$=Mo, W and $X$=S, Se. 
We consider, as representative cases: monolayer compounds (1L),
bilayer compounds (2L), and bulk compounds, in the conventional
2H-$MX_2$ stacking.
Intermediate multilayer systems with $2 < N < \infty$
do not present any qualitative difference with respect to the cases
$N=2$ and bulk.

Lattice parameters in the perfect crystal structure for the bulk
structure are taken from
Ref. \onlinecite{brumme}.
Monolayer  and bilayer compounds are obtained thus by adding a vacuum
region up to 10 \AA\, between the monolayer/bilayer blocks.
We employ the same computational tools as discussed
in Sec. \ref{s-frozen} for monolayer MoS$_2$.
Van der Waals interactions are thus not included.
The explicit inclusion of van der Waals effects can make
the harmonic probability distribution functions $P(h)$
somehow narrower, but we don't expect substantial
differences in the results.

For each compound, we investigate the effects of the quantum
fluctuations of the interplane $X$-$X$ distance within each sandwich
$MX_2$ (see Fig. \ref{f-structure}e-f).
As mentioned, for monolayer systems this corresponds to the $A_1^\prime$
lattice displacement, whereas for bilayer and bulk compounds
this corresponds to the $A_{\rm 1g}$ lattice displacement.\cite{zhang.phonon,phonon.exp}
Following what done for monolayer MoS$_2$,
for each compound we compute
the bandstructure and
the size of each gap
$E_{\Gamma\rm K}$, $E_{\rm KK}$, $E_{\rm KQ}$ as functions of $h$.
Also relevant, due to the interlayer coupling, will appear
the gap $\Gamma$Q between valence states at $\Gamma$ and conduction
states at Q.
A crucial difference between monolayer and multilayer systems with
($N \ge 2$) is the presence, in multilayer systems, of the interlayer
coupling which split of a sizable amount the energy level at $Q$
in the conduction band and at $\Gamma$
in the valence band.\cite{splendiani2010,noi}
In particular, as we are going to see,
the interlayer splitting of the valence band edge at
the $\Gamma$ point is so large to prevent the possibility
that changes of $h$, within the range of
physically alloweed quantum lattice fluctuations,
can induce a transition
of the valence band edge between the K and $\Gamma$ point,
as in the monolayer systems.
In this perspective, unlike in monolayer case,
in multiayer systems (starting from $N=2$)
the valence band edge at $\Gamma$ appears
as a robust feature against quantum lattice fluctuations.
On the other hand, due to the smaller content of the chalcogen $p_z$
orbital,\cite{noi}
the interlayer splitting of the conduction band 
at the $Q$ point is much weaker,
and it does not prevent thus the $h$-induced transition of the
conduction edge band from K to Q.

\begin{figure}[b]
\includegraphics[width=17cm,clip=]{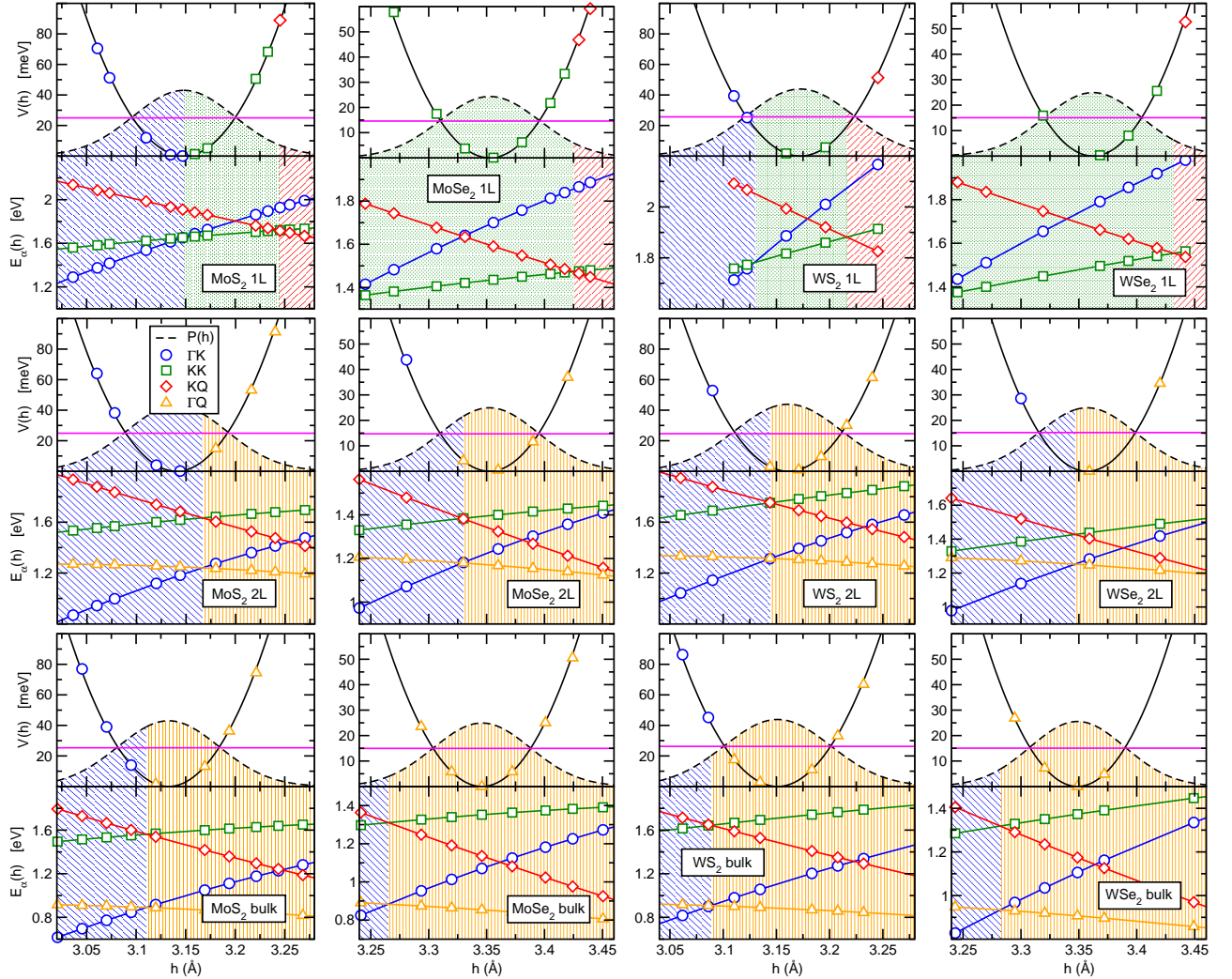}
\caption{
Effects of the quantum lattice fluctuations on the bandgap
for different families of TMDs.
For each compound, in the lower panel we show the evolution
of the different bandgaps $E_\alpha$ as a function of the
interplane $X$-$X$ distance; in the upper panel the frozen
phonon DFT total energy (open symbols), the quadratic fit (solid line),
the energy $\omega_{A_1^\prime/A_{\rm 1g}}/2$ relevant
for lattice quantum fluctuations (horizontal line), and the
probability distribution function $P(h)$ (dashed line).
}
\label{f-panels}
\end{figure}

In addition to the depedence of the band-structure on the parameter $h$,
for each compound
we compute also the frozen phonon energy potential $V(h)$
associated with the $A_1^\prime$/$A_{\rm 1g}$ lattice displacements.
As discussed in the previous section, such joint analysis permits us to
estimate the size of the lattice fluctuations associated
with the zero point motion, and the weights $W_\alpha$
and the average bandgaps $\bar{E}_\alpha$, evaluated
according Eqs. (\ref{weights})-(\ref{averages}).
The results are summarized in Fig. \ref{f-panels} and in Table \ref{t-we}.
\begin{table}
\begin{tabular}{llclcccccccccc}
\hline
\hline
&
& $h_{\rm cl}$ & E$_{\rm cl}$
 & $h_{\rm ZPM}$ & $\omega_{A_1^\prime/A_{\rm 1g}}$  &
$W_{\Gamma{\rm K}}$  & $\bar{E}_{\Gamma{\rm K}}$ &
$W_{\rm KK}$ & $\bar{E}_{\rm KK}$  & 
$W_{\rm KQ}$ & $\bar{E}_{\rm KQ}$  & 
$W_{\Gamma\rm Q}$ & $\bar{E}_{\Gamma\rm Q}$ \\
&
&  (\AA) &
 &  (\AA) &  (meV) &
 &  (eV) &
 &  (eV) & 
 &  (eV) & 
 &  (eV)\\
\hline
MoS$_2$ & (1L) & 3.148 & KK & 0.051 & 50.0 &
0.50 & 1.502 &
0.47 & 1.671 &
0.03 & 1.681 & - & -
\\
MoS$_2$ & (2L) & 3.140 & $\Gamma$K & 0.051 & 49.8 &
0.69 & 1.068 &
- & - &
- & - & 0.31 & 1.217
\\
MoS$_2$&  (bulk) & 3.132 & $\Gamma$Q & 0.051 & 50.6 &
0.33 & 0.787 &
- & - &
- & - & 0.67 & 0.867
\\
\hline
MoSe$_2$ & (1L) & 3.352 & KK & 0.043 & 29.2 &
- & - &
0.95 & 1.427 &
0.05 & 1.441 & - & -
\\
MoSe$_2$ & (2L) & 3.352 & $\Gamma$Q & 0.042 & 29.4 &
0.39 & 1.049 &
- & - &
- & - &
0.61 & 1.146
\\
MoSe$_2$ & (bulk) & 3.346 & $\Gamma$Q & 0.042 & 29.9 &
0.03 & 0.855 &
- & - &
- & - &
0.97 & 0.851
\\
\hline
WS$_2$ & (1L) & 3.172 & KK & 0.050 & 51.3 &
0.20 & 1.690 &
0.61 & 1.831 &
0.19 & 1.828 & - & -
\\
WS$_2$ & (2L) & 3.161 & $\Gamma$Q & 0.050 & 50.2 &
0.37 & 1.203 &
- & - &
- & - &
0.63 & 1.293
\\
WS$_2$ & (bulk) & 3.151 & $\Gamma$Q & 0.050 & 52.6 &
0.11 & 0.835 &
- & - &
- & - &
0.89 & 0.877
\\
\hline
WSe$_2$ & (1L) & 3.362 & KK & 0.042 & 30.1 &
- & - &
0.95 & 1.484 &
0.05 & 1.521 & - & -
\\
WSe$_2$ & (2L) & 3.357 & $\Gamma$Q &0.042 & 30.4 &
0.40 & 1.173 &
- & - &
- & - & 
0.60 & 1.234
\\
WSe$_2$ &  (bulk) & 3.349 & $\Gamma$Q &0.041 & 30.1 &
0.05 & 0.893 &
- & - &
- & - & 
0.95 & 0.903
\\
\hline
\hline
\end{tabular}
\caption{DFT evaluation of the relevant bandgap character
at the classical level as well in the presence of quantum
lattice fluctuations, for different families of TMDs.
For each material, we report here the classical coordinate
$h_{\rm cl}$,
the corresponding value and character of the bandgap $E_{\rm cl}$ at the
classical level, the magnitude of the zero point motion
lattice fluctuations $h_{\rm ZPM}$, the frequency of the relevant
phonon mode, and the weight $W_\alpha$ and average gap $\bar{E}_\alpha$
of the relevant gaps.
}
\label{t-we}
\end{table}

We see that the magnitude of the interplane lattice fluctuations
$h_{\rm ZPM}$ is essentially ruled only by the atomic weight
of the chalcogenide atom $X$=S, Se according to the scaling
$h_{\rm ZPM}\approx 1/\sqrt{M_X}$.
In addition we note that three different bandgap characters
($\Gamma{\rm K}$, KK and KQ) can be
probed
in single-layer $M$S$_2$ compounds, whereas only bandgaps
with KK and KQ character are spanned in the relevant physical range
in single-layer $M$Se$_2$.

As mentioned above, in multilayer systems,
due to the interlayer coupling,
a direct gap KK 
is not possible,
 but two phases with $\Gamma{\rm K}$
and $\Gamma$Q character (respectively for small $h$ and large $h$)
are possible.
A classical analysis, in our GGA calculations, would predict a well-defined
$\Gamma{\rm K}$ character for bilayer MoS$_2$ and
a $\Gamma$Q character for all the other structures,
while both configurations are relevant at a quantum level,
with different weights.
Within this framework one can predict in addition
that the phase $\Gamma$Q will have a stronger weight in the bulk structures
than in the bilayer one.
At the same time, one can expect a stronger character $\Gamma$Q
in the $M$Se$_2$ compounds than
in the corresponding $M$S$_2$ materials.

\section{Direct-indirect gap tuning in in-plane strained TMDs}
\label{IV}

In the previous Section we have shown how the direct/indirect
character of the band structure of monolayer $MX_2$ is highly
sensitive to the out-of-plane interplane $X$-$X$ distance $h$.
We have also shown how the intrinsic quantum lattice fluctuations
of zero point motion are expected to span different direct/indirect
configurations, so that the typical direct character, predicted
by ab-initio calculations in the perfect crystal structure,
should be revised and a coexistence of different direct/indirect gap characters
can be induced by the lattice quantum fluctuations.

The relevance of this analysis can be also investigated
in the reverse scenario, namely in systems where static DFT
calculations would predict an indirect gap and where
a direct-gap component can be triggered in
as effect of the quantum lattice fluctuations.
This possibility is ruled out in bilayer and multilayer
compounds where, as shown in Fig. \ref{f-panels} and Table \ref{t-we},
no significant direct gap can be induced even by the zero point motion
lattice quantum fluctuations.
The investigation appears on the other hand interesting
in monolayer TMDs under in-plane stress where,
at the static lattice level,
a transition from direct to indirect gap
upon applying in-plane strain was discussed
in many works.\cite{scalise1,johari,lu,yun,peelaers,shi,horzum,ghorbani,zhang,scalise2,guzman,wang}

\begin{figure}[b]
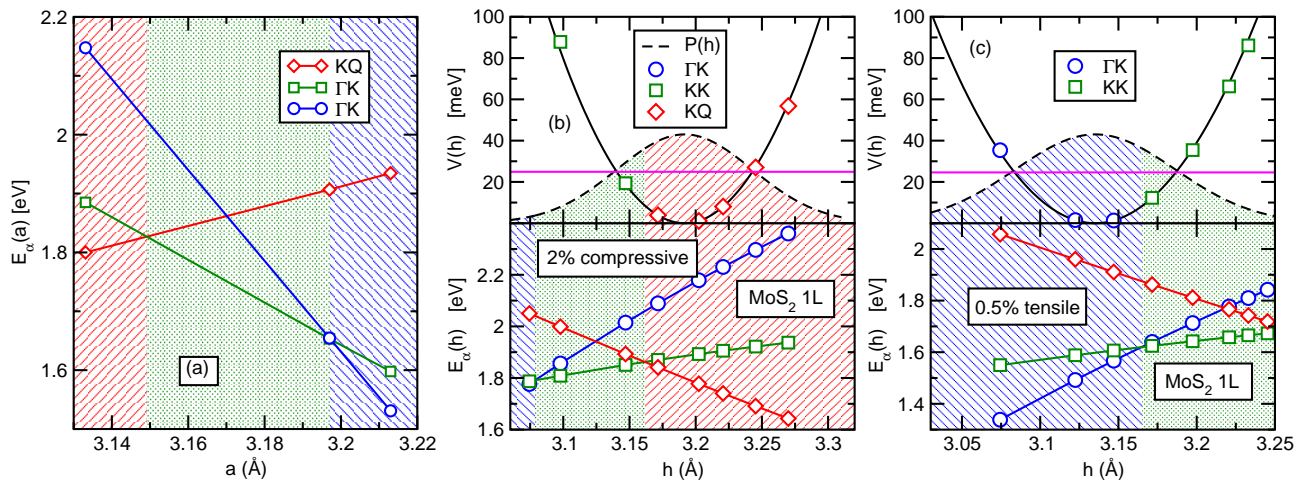

\includegraphics[height=6.3cm,clip=]{f-gap_MoS2_1L_vs_a_3.eps}
\includegraphics[height=6.3cm,clip=]{f-p_MoS2_1L_2c_vs_h.eps}
\includegraphics[height=6.3cm,clip=]{f-p_MoS2_1L_05t_vs_h.eps}
\caption{
(a) Bandgaps $E_\alpha$ at the classical level (i.e. using
$h_{\rm cl}$) for different in-plane lattice constants $a$ of MoS$_2$ monolayer.
The values of $h_{\rm cl}$ for each $a$ are shown
in Table \ref{t-we-strain}. (b)-(c) Evolution of the bandgaps
$E_\alpha$ and of the frozen phonon energy with $h$
for 2 \% compressiove in-plane strain (panel b);
and for 0.5 \% tensile strain (panel c).
}
\label{f-strain}
\end{figure}

To explore this scenario we consider, as representative case,
single-layer MoS$_2$ under biaxial in-plane uniform strain, both tensile and compressive.
In Fig. \ref{f-strain}a we show the phase diagram
as a function of the in-plane lattice constant $a$, computed
in our GGA calculations neglecting quantum fluctuations, i.e.
using the parameter $h_{\rm cl}$ evaluated at the classical level.
As shown in Sec. \ref{mos2.1L} in our GGA calculations (without
dispersion term corrections) the unstrained case $a=3.197$ \AA\,
corresponds to a direct KK gap situation very close to the
change towards an indirect gap with $\Gamma$K character.
We find thus that, at a static level,
a very small tensile strain would be enough to induce a KK
$\rightarrow$ $\Gamma$K transition, whereas
a ~1.5 \% of compressive strain is needed to
induce a KQ gap.
We stress once more that the precise determination of the
such phase diagram , and the location within it of the
unstrained case, might depend on details of the DFT calculations.
The qualitative picture is however robust and general, with
a KQ gap character for compressive strain, a direct KK
bandstructure more or less in the region of absence
of in-plane strain, and a $\Gamma$K bandgap in the tensile strain regime.

Such well-defined bandgap character is however
questioned once intrinsic quantum lattice fluctuations
are taken into account.
This is shown in Fig. \ref{f-strain}b,c where the 
bandgaps $E_\alpha(h)$ 
and the frozen phonon energy $V(h)$ as functions of
the interplane distance $h$ are evaluated.
As done in the previous sections,
the effects of such zero point motion lattice fluctuations are
estimated by using the probability distribution $P(h)$
and the
probability weights $W_\alpha$ and the corresponding
averaged bandgap $\bar{E}_\alpha$
are reported in Table \ref{t-we-strain}.
For sake of comparison, we report in the same table
also the
probability weights $W_\alpha$ and the corresponding
averaged bandgap $\bar{E}_\alpha$ for single-layer MoS$_2$
without strain.
\begin{table}
\begin{tabular}{lcclcccccccc}
\hline
\hline
\multicolumn{12}{c}{single-layer MoS$_2$ under strain} 
\\
\hline
\hline
strain & $a$ 
& $h_{\rm cl}$ & E$_{\rm cl}$
 & $h_{\rm ZPM}$ & $\omega_{A_1^\prime/A_{\rm 1g}}$  &
$W_{\Gamma{\rm K}}$  & $\bar{E}_{\Gamma{\rm K}}$ &
$W_{\rm KK}$ & $\bar{E}_{\rm KK}$  & 
$W_{\rm KQ}$ & $\bar{E}_{\rm KQ}$   \\
\% & (\AA)
&  (\AA) &
 &  (\AA) &  (meV) &
 &  (eV) &
 &  (eV) & 
 &  (eV)\\
\hline
-2.0  & 3.133 & 3.1916 & KQ & 0.051 & 49.8 &
0.01 & 1.748 &
0.27 & 1.837 &
0.72 & 1.753 
\\
0.0  & 3.197 & 3.1481 & KK & 0.051 & 50.0 &
0.50 & 1.502 &
0.47 & 1.671 &
0.03 & 1.681 
\\
+0.5  & 3.213 & 3.1351 & $\Gamma$K & 0.051 & 49.2 &
0.72 & 1.456 &
0.28 & 1.640 &
- & - 
\\
\hline
\hline
\end{tabular}
\caption{Zero point motion $h_{\rm ZPM}$,
probability weights $W_\alpha$, and
averaged bandgap $\bar{E}_\alpha$ for single-layer MoS$_2$
under strain.
}
\label{t-we-strain}
\end{table}
From this analysis, it appears that the net direct/indirect gap transition upon strain
predicted by standard ab-initio calculation neglecting the quantum
lattice fluctuations must be regarded more as a smooth continuous
crossover
between the relative phase spaces available for each bandgap
character.
Indeed, as we are going to discuss in more details, the very presence
of a sizeable phase space for the direct KK gap configuration can hamper
the possibility of observing such crossover in photoluminescence
optics.
Alternative paths for observing such direct-indirect gap crossover are
discussed in the next section.

\section{Discussion}
\label{s:bo}

In the previous Sections, we have revised the concept of
direct/indirect gap character in layered transition
metal-dichalcogenides $MX_2$, which is usually discussed in literature
on the basis of  a well-defined band-structure,
usually evaluated in first-principle calculations
in the perfect crystal lattice, neglecting thus the quantum lattice dynamics.
This assumption is quite justified in conventional semiconductors
where there is a unique valence/conduction band edge,
whose energies can be affected by the lattice dynamics but not their
location in the Brillouin zone.
From this perspective, the crucial novelty of TMDs is the presence
of secondary band edges, both in the valence and conduction sectors,
which lie only few tenths of meV above/below the true band edges.
Within this context, we have shown how a direct/indirect bandgap transition
can occur as a function of the inter-plane distance $h$ between the
two $X$ planes, and how intrinsic zero point motion
quantum lattice fluctuations of this distance
can effectively span different bandgap topologies, from a direct one at
K-K, to indirect ones of $\Gamma$-K or K-Q character.
The relevance of the different band-structure probed by the lattice
dynamics has thus been quantified in terms of appropriate
spectral weight $W_\alpha$ and corresponding average bandgap energies
$\bar{E}_\alpha$.
On this basis we can conclude that it is not possible to discuss
electronic particle-hole excitations in terms of a well-defined direct/indirect
bandgap paradigm.

A compelling analysis of the implications of these effects
with respect to the physical properties observed in optics appears
highy desiderable ,
but it requires to be formally investigated in a context which goes
beyond the standard Born-Oppenheimer scheme.\cite{bcp,cp}
This is a formidable task which cannot be addressed here.

On this regards, it is worth mentioning that a promising approach,
merging ab-initio and quantum field theory calculations,
has been recently developed,\cite{cannuccia1,cannuccia2,cannuccia3,ponce}
starting from the seminal ideas of M. Cardona,\cite{ct,ah,ac}
for evaluating the effects of the zero point quantum lattice
fluctuations on the band structure.
The basic idea of such approach stems from a careful identification
of the Feynman's diagrams associated with the zero point motion.
However, although powerful, such theory is at the present mainly focused
on the {\em single-particle} excitations, namely the one-particle
Green's function and the one-particle self-energy.
These quantities, evaluated in the presence of quantum lattice
fluctuations, are thus used as the basic ingredients to define
an renormalized effective band structure.
It is interesting to stress that this approach permits not only to
define (as many other approaches)
dynamically renormalized quasi-particle excitation energies,
but also to associate to them a {\em finite} lifetime/linewidth.
The presence of a finite linewidth would smear the band structure
(see for instance Fig. 2 in Ref. \cite{cannuccia1},
in similar way as the dependence of the band-structure
on the parameter $h$ would give a band smearing
in Fig. \ref{f-bandsMoS2_h}a.\cite{bcp}
The {\em possibility} of a direct/indirect gap crossover
would be detected within the context
of Refs. \onlinecite{cannuccia1,cannuccia2,cannuccia3,ponce}
when the smearing of a secondary band edges
is larger than the distance with the primary band edge.
Note however that, in such theory, it is not possible to investigate the
coherent shift of the band edges at the $\Gamma$, K, Q points
upon the lattice dynamics of the quantity $h$,
as efficiently done in Fig. \ref{f-bandsMoS2_h}a
The actual direct/indirect gap transition cannot be thus effectively assessed.

In this sense, the present analysis is complementary of the approach
of Refs. \onlinecite{cannuccia1,cannuccia2,cannuccia3,ponce},
permitting to identify the effective direct/indirect gap transition.
Merging the two approaches would be highly desirable.
In particular, the main challenge in this context is the prediction
of observable nonadiabatic effects in the optical probes.
A possible progress along this line would be the generalization
of the Cardona theory to the particle-hole response function,
in a conserving approach consistent with the one-particle
resummation in the self-energy.

Although a compelling theory of the nonadiabatic transition
between direct and indirect bandgap configurations
in these materials is still lacking, few qualitative considerations
about the possible effects on physical observables
can at the moment be drawn.
For sake of simplicity, we will consider three representative regimes:
$i$) a system where classical ab-initio calculations would predict
a direct gap but where quantum lattice fluctuations suggest
a sizable relevance of indirect gap configurations;
$ii$) a system where classical ab-initio calculations would predict
an indirect gap but where quantum lattice fluctuations suggest
a sizable relevance of direct gap configurations.
$iii$) a system where classical ab-initio calculations would predict
an indirect gap but where quantum lattice fluctuations suggest
a sizable relevance of an indirect gap of different kind.

Far from being hypothetical cases, these examples
represent different possible regimes that have been so far intensively studied
for theoretical and application purposes:
$i$) monolayer dichalcogenides $MX_2$ in the absence of in-plane
strain;
$ii$) monolayer dichalcogenides $MX_2$ in the presence of a large enough
in-plane strain inducing (at a classical level) as indirect gap band structure;
$iii$) multilayer dichalcogenides $MX_2$.

Investigating the effects of quantum lattice fluctuations in the
regime ($i$) is probably the most crucial one, since
the direct bandgap character has been claimed to be observed
in many monolayer compounds (e.g. MS$_2$, WS$_2$, \ldots)
by means of photoluminescence probes.
On this regards, it is worth to stress that the presence of a sizable
component of a indirect-gap band-structure
is overall compatible and not at odds with such phenomenology since
the strong intensity of the direct-gap photoemission is expected
to be dominant with respect to other indirect-gap electronic
configurations probed by the lattice quantum fluctuations.
Similar considerations holds true for many of the optical probes
related to the A and B excitons,
taking into account that the band structure (and hence the
optical transitions) at K point are relatively insensitive
to the lattice dynamics (see Fig. \ref{f-bandsMoS2_h}a for example).
On the other hand the spectral features associated with the
exciton C are expected to be highly affected by the quantum lattice
fluctuations. This might account for large broadening and for the
complex structure of the C exciton as observed in
optical probes.\cite{qiu,kozawa,yu.c,wang.c}
In similar way, traces of indirect gaps can be possibly
observed in photoluminescence at different energies than
the direct gap.\cite{WSe2mono-direct}
The presence of a conduction band-edge minimum at Q probed
by the lattice fluctuations can have also remarkable consequences
on transport properties, as explored by photoconductivity or
in field-effect geometry of chemical doped systems.

Relevant effects induced by the quantum lattice fluctuations can be
expected in the case ($ii$), representative for instance of
monolayer systems under in-plane strain.
Here ab-initio calculations with frozen lattice coordinates
would classify the system as an indirect bandgap semiconductor,
whereas we would predict that a
sizable direct-gap band-structure is dynamically probed
by the quantum fluctuations.
On a qualitative ground, we would expect that
the presence of an even minority probability
of a direct-gap configuration would be the dominant feature
in a photoluminescence measurement, possibly with reduced
intensity, so that photoluminescence experiments would
not be able to distinguish this case from a pure
direct gap case.
Nevertheless, at the same time transport properties,
as in case ($i$),
are expected to be dominated by the indirect-gap character.
The crossed analysis of photoluminescence and transport experiments
can provide thus a route to characterize this regime.
Further measurements along this line are thus encouraged.

In case ($iii$) the location of the valence band-edge at $\Gamma$
appears robust against the presence of quantum lattice fluctuations.
Such quantum fluctuactions however can drive a dynamical tuning
between a minimum conduction band-edge at K or Q.
This kind of scenario is probably the most difficult to assess
experimentally. Suitable ways to detect this situation rely probably
in the possibility of accurate polarized optical probes where
the different chiral content (and hence sensitivity
to polarized light) of the minima in K and Q can be investigated.
Physical consequences of this scenario can be also relevant
in understanding and characterizing the observed superconducing phase
in heavily electron-doped  multilayered MoS$_2$ and
other transition-metal
dichancogenides.\cite{ye.sc,roldan.sc,yu.sc,shi.sc,jo,costanzo,saito}

\acknowledgments

L.O. acknowledges support from CINECA through the
IsC35 TDM01 ISCRA project.
E.C. acknowledges financial support from the
Italian MIUR-PRIN Project 2015WTW7J3.

\end{document}